\documentclass[journal=jacsat,manuscript=article]{achemso}

\usepackage[version=3]{mhchem} 

\author{J. C. Garcia}
\author{J. F. Justo}
\email{jjusto@lme.usp.br}
\affiliation[Escola Polit\'ecnica, Universidade de S\~ao Paulo,
CP 61548, CEP 05424-970, S\~ao Paulo, SP, Brazil]
{Universidade de S\~ao Paulo, S\~ao Paulo, SP, Brazil}
\author{W. V. M. Machado}
\author{L. V. C. Assali}
\affiliation[Instituto de F\'{\i}sica,Universidade de S\~ao Paulo,
CP 66318, CEP 05315-970, S\~ao Paulo, SP, Brazil]
{}

\title[adamantane]
{Structural, electronic and vibrational properties of amino-adamantane 
and rimantadine isomers}

\begin{document}

\begin{abstract}  
We performed a first principles total energy investigation on the structural, 
electronic, and vibrational properties of adamantane molecules, functionalized 
with amine and ethanamine groups. We computed the vibrational signatures of 
amantadine and rimantadine isomers with the functional groups bonded to 
different carbon sites. By comparing our results with recent 
infrared and Raman spectroscopic data, we discuss the  possible
presence of different isomers in experimental samples.    
\end{abstract}


\section{Introduction}  
\label{sec1}  

Diamondoids, also known as molecular diamond, are nanostructured 
forms containing diamond-like carbon cages passivated with hydrogen 
atoms. Although they have been known for a long time, only the recent 
successes in separating them in different sizes and shapes \cite{dahl} 
allowed to explore several of their potentialities. Diamondoids have been 
considered for widespread technological applications, such as electronic 
devices \cite{fp,yywang},  nano-electro-mechanical 
systems \cite{garcia,garcia2}, and pharmaceuticals \cite{huang,stouffer}. 
Over the last few years, pharmaceutical applications have flourished, 
since functionalization of adamantane with amine groups leads to 
molecules that can be used to treat several diseases, such as influenza A, 
depression, and even some cases of parkinsonism.

Functionalization of adamantane ($\rm C_{10}H_{16}$) with the amine 
group ($\rm -NH_2$) leads to amantadine ($\rm C_{10}H_{17}N$), also 
known as amino-adamantane. Functionalization with the ethanamine group 
($\rm -CHCH_3NH_2$) leads to rimantadine ($\rm C_{12}H_{21}N$), also known 
as adamantyl ethanamine. Those two chemical compounds are used to treat 
the influenza A virus, and are already present in several commercial drugs. 
Those molecules act over the influenza virus, by binding into the proton 
channels of the M2 viral protein and blocking protonation, which is a 
fundamental step for virus replication \cite{schnell}. However, the 
microscopic mechanisms of such protein binding are still subject of great 
controversy \cite{khurana}. This virus strain has recently acquired 
considerable resistance to amantadine and rimantadine 
treatments \cite{stouffer,pielak}. Therefore, it became fundamental 
to understand the microscopic mechanisms of the interactions between 
those molecules and the virus membranes, in order to provide enough 
information to address the upcoming challenges to treat influenza.

\begin{figure}
\centerline{\includegraphics[width=160mm]{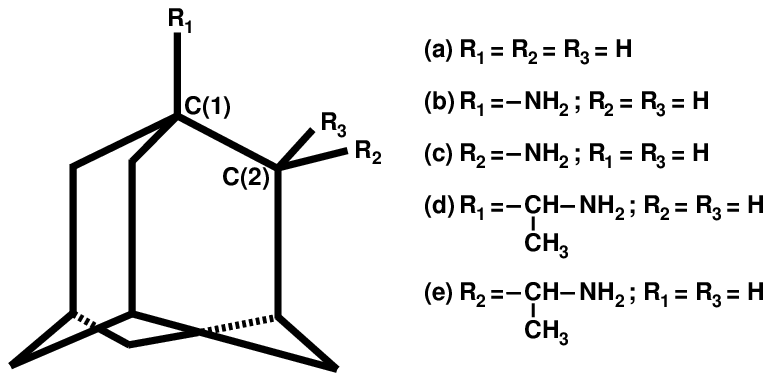}}
\caption{Schematic representation of the molecular
forms considered in this investigations: (a) adamantane, 
(b) 1--amino-adamantane (1--amantadine), (c) 2--amino-adamantane 
(2--amantadine), (d) 1--rimantadine, and (e) 2--rimantadine. 
Hydrogen atoms are not represented in the figure, except those
present in the radicals (R$_1$, R$_2$, and R$_3$).}
\label{fig1}  
\end{figure}

Adamantane consists of a single diamond-like carbon cage and carries 
two types of carbon atoms, represented schematically in Fig. \ref{fig1}, 
labeled C(1) and C(2). A C(1) is bound to three C(2)'s and one H atom, 
while a C(2) is bound to two C(1)'s and two H's. The amine group 
(-NH$_{2}$) could bind to either C(1) or C(2), forming respectively 
1--amino-adamantane (here labeled 1--amantadine)  or   
2--amino-adamantane (here labeled 2--amantadine) isomers, 
schematically represented in Figs. \ref{fig1}b and \ref{fig1}c. 
The 1-- and 2--rimantadine isomers, as represented respectively in 
Figs. \ref{fig1}d and \ref{fig1}e, could be formed by binding the 
ethanamine ($\rm -CHCH_3NH_2$) group to either C(1) or C(2) sites.  
Although the respective isomers are chemically equivalent, their structural 
and conformational differences may be relevant when interacting with the M2 
virus proton channels. As it has been recently pointed out, the conformation 
of the amantadine molecules within those narrow proton channels determines 
its effectiveness in blocking virus protonation \cite{stouffer}. In fact, 
some studies have already discussed the relevance of isomers in treating 
influenza \cite{zoidis}. The signatures of different isomeric molecules 
could be identified by comparing the respective vibrational modes coming 
from infrared (IR) and Raman spectra. Here, we used first principles total 
energy calculations, based on the Gaussian simulation package, to 
investigate the structural, electronic, and vibrational properties of 
amantadine and rimantadine isomers. Our results were compared to available 
experimental data, and suggested the presence of different isomeric forms 
in samples.

\section{Methodology}  
 \label{sec2}
 
The stable geometries, electronic structures, and vibrational modes of 
amantadine and rimantadine isomers were investigated using the Gaussian98 
package \cite{gaussian}. We considered a hybrid density functional theory 
(DFT), combining the Becke's three-parameter hybrid functional \cite{becke} 
and the correlation one of Lee, Yang, and Parr \cite{lyp} (B3LYP). The 
electronic states were described by a 6-311++G** Pople basis set, which 
consists of a split valence triple-zeta basis augmented with diffuse (+) 
and polarized (*) functions. In all calculations, self-consistent cycles 
were performed until differences in total energies were lower than 
10$^{-6}$ Ha between two iterations. The geometry optimizations were 
performed using the Berny's algorithm \cite{peng} with no symmetry 
constrains. A converged geometry was obtained when the total force in any 
atom was lower than 10$^{-3}$ Ha/a.u. This theoretical framework, based on 
the B3LYP/6-311++G** combination, has been widely used to explore the 
structural and vibrational properties of several organic 
molecules \cite{rama,merrick,bas1,kurt,bas2}, providing a good agreement 
with respective experimental data. Therefore, it is expected that it would also 
provide an appropriate description of amantadine and rimantadine properties.

We used the adamantane molecule as a control test \cite{ricky}, and checked 
the reliability of this methodology in describing the structural and 
vibrational properties of this molecule with respect to other models and 
basis sets. In terms of the exchange and correlation potentials, a comparison 
for adamantane, considering the DFT/B3LYP and the MP2  models \cite{mp} and a 
standard 6-31G*, shows that results on interatomic distances 
(within 1\% of accuracy), angles (within 1\% of accuracy), and 
vibrational modes (within 2\% of accuracy) are consistent with each 
other \cite{jensen}. However, the DFT/B3LYP functional provides results which 
are considerably better than those using a Hartree-Fock level approximation, 
when compared to experimental data \cite{jensen}.

In terms of the basis set, we compared results for adamantane, using the 
6-311++G** and 6-31+G* sets, and found that interatomic distances and
angles differ by less than 0.2\% and vibrational frequencies 
(for both IR and Raman modes) differ by less than 1\% for most of the 
modes, except for the CH and CH$_2$ stretching modes, that differs by about 2\%.  
Additionally, as it has been recently shown, convergence for structural and 
electronic properties of adamantane,  in terms of number of orbitals in 
the basis set, is achieved for the 6-311++G** set \cite{marsusi}. In the 
following section, a methodological certification is presented for the 
adamantane properties, by comparing our results with available experimental 
data.

\section{Properties of adamantane: control test}
\label{sec3}

Using the theoretical framework described in the previous section, the 
optimized adamantane molecule presented interatomic distances and bond angles
 which are in excellent agreement with the experimental data (within 1\% of 
accuracy) for the molecule in the gas phase \cite{hr}. Additionally, our 
results are in excellent agreement with those from investigations 
using different theoretical approaches \cite{garcia,rama,ricky,mbb}.

The energy gap of this molecule ($\rm \Delta E_{HL}$), the difference in energy 
between the highest occupied molecular orbital (HOMO) and the lowest 
unoccupied molecular orbital (LUMO), is 7.22 eV, consistent with the value 
from a different investigation \cite{marsusi}. Additionally, this value 
should be compared to optical gap 6.49 eV from recent experimental 
measurements \cite{landt} and to the value of 7.62 eV using a different 
theoretical model \cite{mbb}. The HOMO has a bonding character, and its 
spatial charge distribution, presented in Fig. \ref{fig2}a, is associated 
with the interstitial region between the C(1) atoms and their respective 
H neighbors. On the other hand, the LUMO has an anti-bonding 
character, associated with the C(1)-C(2) interaction, with the charge 
distribution in the back-bond of C(2) atoms.

As a stringent test on the reliability of this methodology, we computed the IR 
and Raman spectra of adamantane and compared with available data, as shown in 
figure \ref{fig3}. The vibrational spectra of adamantane consists essentially 
of five distinct regions \cite{filik1}: 400-700 cm$^{-1}$ (CCC bending modes), 
700-1000 cm$^{-1}$ (CC stretching modes), 
1000-1400 cm$^{-1}$ (CCH$_2$ rocking, twisting, 
and wagging modes),  1400-1600 cm$^{-1}$ (CH$_2$ scissor 
modes), and 2700-3000 cm$^{-1}$ (CH and CH$_2$ 
stretching modes). The figure indicates that theoretical results for most of 
the modes are within 1\% of experimental data for both IR and Raman spectra. 
However, for the CH and CH$_2$ stretching modes, theoretical values 
underestimate frequencies by about 3\%, compared to experimental data. Such 
theoretical limitation, in treating high frequency vibrational modes, is well 
documented in the literature \cite{ricky}. Those stretching modes are associated 
with vibrations involving variations in the C-H interatomic distances in the 
external molecular bonds, being very sensitive to the molecular 
environment. It turns out that theoretical results are related to molecules 
in the gas phase, while experimental data are related to solid or solution 
phases, in which those external C-H bonds interact with the environment. 
For all the other vibrational modes, results are insensitive to the 
molecular environment, since the respective bonds are internal to the molecule. 
It has also been argued in the literature that such 
discrepancy in high frequency modes might come from strong anharmonic interactions 
between closely spaced modes of equal symmetry \cite{oomens}. We stress that 
increasing the number of orbitals in the basis set does not improve the 
description of modes in that high frequency range, consistent with
other investigations \cite{oomens}.

\section{Properties of amantadine isomers}  
\label{sec4}
    
Functionalization of adamantane with the amine group, to form  1-- or 
2--amantadine, causes negligible structural changes in interatomic 
distances and angles (in atoms far from the functionalized site)
by less than 0.2\%, in comparison with those of the original adamantane 
molecule. Near the functional group, the C(1)-C(2) interatomic distances 
are slightly affected, but important changes are observed in 
bond angles. In 1--amantadine, two C(2)-C(1)-N angles are 
108.7$^{\circ}$, while the third one is 113.7$^{\circ}$. In 
2--amantadine, those angles are 110.9$^{\circ}$ and 
116.0$^{\circ}$, respectively. In both isomers, C-N and N-H distances 
are 1.469 and 1.016 \AA, respectively. The introduction of the amine 
group in adamantane causes important changes on the electronic structure 
of the original molecule. The resulting $\rm \Delta E_{HL}$ for 1-- and 
2--amantadine is 6.03 and 5.99 eV, respectively. As shown in figure 
\ref{fig2}b and \ref{fig2}c, the HOMO of both isomers are mostly related 
to the nitrogen 2p non-bonding lone-pair. On the other hand, their HOMO-1 
resemble the HOMO of adamantane. The LUMO for both isomers resembles the 
one of adamantane, including now some small charge distribution in the 
back-bond of their nitrogen atoms. Therefore, the reduction in 
$\rm \Delta E_{HL}$, from adamantane to 1-- or 2-- amantadine, may be 
understood as resulting from the introduction of a nitrogen 2p non-bonding 
orbital above the HOMO of adamantane, while the LUMO is essentially unaffected 
by the amine incorporation. Those results also indicate that the 
isomers have a highly reactive site in the amine group, which is related
to the nitrogen lone-pair.

In terms of stability, we obtained that the total energy of 2--amantadine 
molecule is only 0.17 eV higher than that of 1--amantadine, which is in
excellent agreement with the experimental value for the difference 
in heat of formation of 0.16 eV at room temperature \cite{clark}.    
Therefore, although samples should contain a prevailing concentration of 
1--amantadine, the presence of 2--amantadine should not be negligible. In order 
to address this point, and confirm the presence of different isomers in 
experimental samples, we investigated the vibrational signatures of each isomer 
from IR and Raman spectroscopic data, and compared to available data.

Figure \ref{fig4} shows the theoretical IR and Raman vibrational 
spectra of 1-- and 2--amantadine, and the respective experimental spectra of 
amantadine samples (in condensed phases) \cite{nist,rivas}. A major effect of 
amine incorporation in adamantane, for both isomers, is a symmetry lowering that 
causes important splitting in most of the vibrational modes. Additionally, the 
spectra present new nitrogen-related vibrational modes.

In the IR spectra, it is possible to clearly identify vibrational signatures 
of each isomer. Table \ref{tab1} presents the assignments of the important 
vibrational modes of those isomers, according to figure \ref{fig4}. The CH and
CH$_2$ stretching modes appear in the 3000 cm$^{-1}$ region, and there are 
similarities 
in the spectra for 1-- or 2--amantadine, such that it cannot be used to identify 
a signature of either isomer in the experimental data, that appears in the 
2853-2932 cm$^{-1}$ range. The CH$_2$ scissor modes appear on 1493 
and 1500 cm$^{-1}$, respectively for 1-- and 2--amantadine, while experimental 
spectra show a single strong peak in 1458 cm$^{-1}$.
The CC stretching modes appear as a set of several peaks in the 806-1042 and 
748-1069 cm$^{-1}$ range for 1-- and 2--amantadine, respectively. Experimental 
IR spectra shows a set of several peaks in that region. The functional group 
introduces several nitrogen related vibrational modes. The NH$_2$ scissor mode 
appears as a single peak on 1652 and 1657 cm$^{-1}$ respectively for 1-- and 
2--amantadine, while experimental data shows two peaks in that region, which 
could be a signature of the presence of different isomers in the samples. 
The NH$_2$ torsion mode appears as a pair of peaks on 245 and 281 cm$^{-1}$ 
for 1--amantadine and as a single peak on 212 cm$^{-1}$ for 2--amantadine. 
Unfortunately experimental data does not 
cover such frequency region to allow a proper identification \cite{nist}. The CN 
stretching mode appears as a set of two peaks (on 1113 and 1162 cm$^{-1}$) 
for 1--amantadine, but as a single peak (on 1121 cm$^{-1}$) for 2--amantadine. 
In that region, experimental data present a set of several strong peaks 
(on 1103, 1153, and 1193 cm$^{-1}$). The poor resolution of experimental IR data, 
due to measurements in condensed phases that causes broadening on the peaks, 
compromises a clear comparison with theoretical spectra, although results indicate 
a prevailing concentration of 1--amantadine.

The Raman spectroscopy provides results with good frequency resolution, such that 
a direct assignment of modes and a comparison with experimental data \cite{rivas} 
could be easier than for IR spectroscopy. Figure \ref{fig4} shows the theoretical 
Raman spectra of 1-- and 2--amantadine, while table \ref{tab1} presents the most 
relevant vibrational modes. The NH$_2$ stretching modes, which are inactive by 
IR spectroscopy, show clear peaks in the Raman spectra. The symmetric and 
antisymmetric modes are on 3473 and 3551 cm$^{-1}$ for 1--amantadine and on 3500 
and 3579 cm$^{-1}$ for 2--amantadine. The splitting between symmetric and 
antisymmetric modes is about 80 cm$^{-1}$ for any isomer. On the other hand, there 
is a shift of 30 cm$^{-1}$ in the respective lines for different isomers. This 
could, in principle, serve as a signature to identify the presence of different 
isomers on experimental samples. Unfortunately, available experimental 
data does not cover such frequency range  \cite{rivas}. 
The CC stretching modes appear as a set of several peaks for 1--amantadine, with 
two strong ones on 715 and 775 cm$^{-1}$. For 2--amantadine, there are also 
several peaks, but only one strong on 748 cm$^{-1}$. Experimental data shows two 
strong peaks on 718 and 781 cm$^{-1}$, indicating a prevailing presence of 
1--amantadine in samples. The 1000-1100 cm$^{-1}$ is an 
interesting frequency range to compare results of different isomers. 
Our results show no lines for 1--amantadine but several ones for 
2--amantadine. The presence of some lines on experimental spectra may
indicate traces of 2--amantadine on samples.

\section{Properties of rimantadine isomers}  
\label{sec5}
 
Functionalization of adamantane with the ethanamine group, to form  1-- or 
2--rimantadine, causes negligible structural changes in interatomic distances 
and angles, as compared to  original adamantane molecule. Such changes are 
even smaller than the ones for amine functionalization, discussed in 
section \ref{sec4}. In both isomers, C-N and N-H distances are 1.477 and 
1.016 \AA, respectively. The introduction of the ethanamine group in
adamantane causes important changes on the electronic structure of the
original adamantane molecule. The resulting $\rm \Delta E_{HL}$ for 1-- and 
2--rimantadine are 6.12 and 6.16 eV, respectively. As shown in figure 
\ref{fig2}d and \ref{fig2}e, the HOMO of both isomers are related to the 
nitrogen 2p non-bonding lone-pair, equivalently to those for amantadine molecules. 
The LUMO for both isomers resembles the one of adamantane, including now some 
charge distribution in the back-bond of nitrogen and carbon atoms of 
the functional group. Therefore, the reduction in $\rm \Delta E_{HL}$, from 
adamantane to rimantadine, can be understood equivalently to the case of
amantadine. The results also indicate that the isomers have a highly reactive 
site in the ethanamine group, which is related to the nitrogen lone-pair.

In terms of stability, the total energy of 2--rimantadine molecule is only 
0.23 eV higher than that of 1--rimantadine. Equivalently to the case of 
adamantane, there should be a prevailing concentration of 1--rimantadine in 
experimental samples, but the presence of 2--rimantadine should not be 
negligible. IR and Raman spectroscopy could address this point. Figure \ref{fig5} 
shows the theoretical IR and Raman vibrational spectra of 1-- and
2--rimantadine, and the respective experimental data of rimantadine 
samples (condensed phase of hydrochloride rimantadine) \cite{sigma}.

Figure \ref{fig5} shows the theoretical and experimental IR spectra of 
rimantadine isomers. In the high frequency, there are several CH and CH$_2$ stretching 
modes, which are mixed with the CH$_3$ stretching mode, coming from the functional 
group. That wide band for both isomers, in the 3000 cm$^{-1}$ region, includes 
a tail in the high frequency, which is related to the CH$_3$ group, and is fully 
consistent with experimental data \cite{sigma}. Other vibrational modes 
of rimantadine, described in 
table \ref{tab1}, show similarities with the respective ones of amantadine.
The CH$_2$ plus CH$_3$ rocking
modes appear as a single peak in 1--rimantadine (on 1499 cm$^{-1}$) 
and as a set of two peaks in 2--rimantadine (on 1491 and 1508 cm$^{-1}$). The 
experimental spectra shows a single line on 1519 cm$^{-1}$, suggesting a prevailing 
concentration of 1--rimantadine in samples. 
The NH$_2$ wagging mode appears as a set of two peaks 
in 1--rimantadine (on 842 and 868 cm$^{-1}$) and as a single peak in
2--rimantadine (on 861 cm$^{-1}$), while experimental spectra shows several
peaks in that frequency region. Finally, the NH$_2$ torsion mode appears 
as a set of two peaks on 233 and 250 cm$^{-1}$ for 1--rimantadine
and as a single peak on 215 cm$^{-1}$ for 2--rimantadine, but there
is no experimental data in this frequency region.

Figure \ref{fig5} shows the theoretical and experimental Raman spectra of 
rimantadine isomers \cite{sigma}. Most of the vibrational modes resemble
the ones for amantadine, as described in table \ref{tab1}. 
The ethanamine group introduces CH$_3$ stretching modes, which mixed with the 
CH and CH$_2$  ones, introducing a tail in  
in the high frequency region. The theoretical spectra shows clear NH$_2$ 
symmetric and antisymmetric stretching modes,
in the high frequency region, in  1--rimantadine (on 3491 and 3568 cm$^{-1}$)  
and 2--rimantadine (on 3485 and 3561 cm$^{-1}$). Although such NH$_2$ modes
could be used to identify signatures of either isomers, experimental data does
not appear to cover such frequency region.

\section{Summary}  
\label{sec6}
  
In summary, we carried a theoretical investigation on the structural, electronic, 
and vibrational properties of two isomeric forms of amantadine and rimantadine 
molecules. We find that, for all amantadine and rimantadine isomers, the charge 
distribution of HOMO is always associated with the nitrogen lone pair, which is also 
the most chemically reactive site in those molecules. Therefore, the mechanism of
interaction between those molecules with the viral proton channels is likely 
associated with the protonated nitrogen lone pair. Although the HOMO in different 
isomers have the same nitrogen character, their spatial conformations are different, 
which would cause  different electronic affinity, and consequently, different strength 
in interactions with proton channel walls.

We additionally found small difference in total energies between isomeric forms of 
amantadine and rimantadine, which suggest that both isomers could be present in
samples. The IR and Raman vibrational spectra of those isomers 
were discussed in the context of the changes in spectra of adamantane caused by the 
amine and ethanamine incorporation. The comparison of theoretical and experimental 
spectra suggested a prevailing presence of 1--amantadine and 1--rimantadine molecules 
in experimental samples, although the presence of their respective isomers 
cannot be neglected. Our assignment of all vibrational modes of isomeric forms 
can be useful in future explorations of those molecules.
 
\vspace{0.5cm}

\textbf{Acknowledgments}  
  
The authors acknowledge partial support from Brazilian agencies CNPq and
FAPESP.

\newpage


\begin{table}[htpb]
\caption{Theoretical vibrational (IR and Raman) modes of 
1--amantadine, 2--amantadine, 1--rimantadine, and 
2--rimantadine isolated molecules.
Frequencies are given in cm$^{-1}$.}
\vspace{0.3cm}
\begin{center}
\begin{tabular}{lcccc}
\hline \hline
 & \multicolumn{4}{c} {IR} \\
~mode~~         & ~1--amantadine~ & ~2-amantadine~ & ~1--rimantadine~ & ~2-rimantadine~ \\
\hline
$\nu$(CH$_2$)    & 3004-3056 & 3008-3042 & 3009-3051 & 3001-3047  \\
$\nu$(CH$_3$)    & ---       & ---       & 3081+3095 & 3070+3106  \\
$\delta$(CH$_2$) &  1493     &  1500     & 1499       & 1491+1508 \\ 
$\nu$(CC)        &  806-1042 & 748-1069  & 816-1047   & 824-1052  \\
$\delta$(NH$_2$) & 1652      &   1657    &   1656     &  1657     \\
$\omega$(NH$_2$) &  853      &   851     &  842+868   &   861     \\     
$\tau$(NH$_2$)   &  245+281  &   212     &  233+250   &   215     \\
$\nu$(CN)        & 1113+1162 &   1121    & 1072+1094  & 1068+1117  \\
\hline
 & \multicolumn{4}{c} {Raman} \\
\hline
$\nu$(CH$_2$)    & 3005-3056 & 2006-3044 & 3002-3043  & 3001-3033 \\
$\nu$(CH$_3$)    & ---       & ---       & 3081+3095  & 3070+3106 \\
$\delta$(CH$_2$) &  1479     & 1485-1517 & 1484-1522  & 1485-1517 \\ 
t(CH$_2$)        & 1223      &  1237     & 1212+1232  & 1226+1242 \\ 
$\nu$(CC)        &  715-998  & 748-986   & 701-991    & 762-993   \\
$\nu$(NH$_2$)    & 3473+3551 & 3500+3579 &  3491+3568 & 3485+3561 \\
$\delta$(NH$_2$) &  1652     &   1657    &   1656     &  1657     \\
\hline \hline
\end{tabular}
\end{center}
\label{tab1}
\end{table}


\pagebreak   
\begin{small}  
\begin{figure}  
\centerline{\includegraphics[width=125mm]{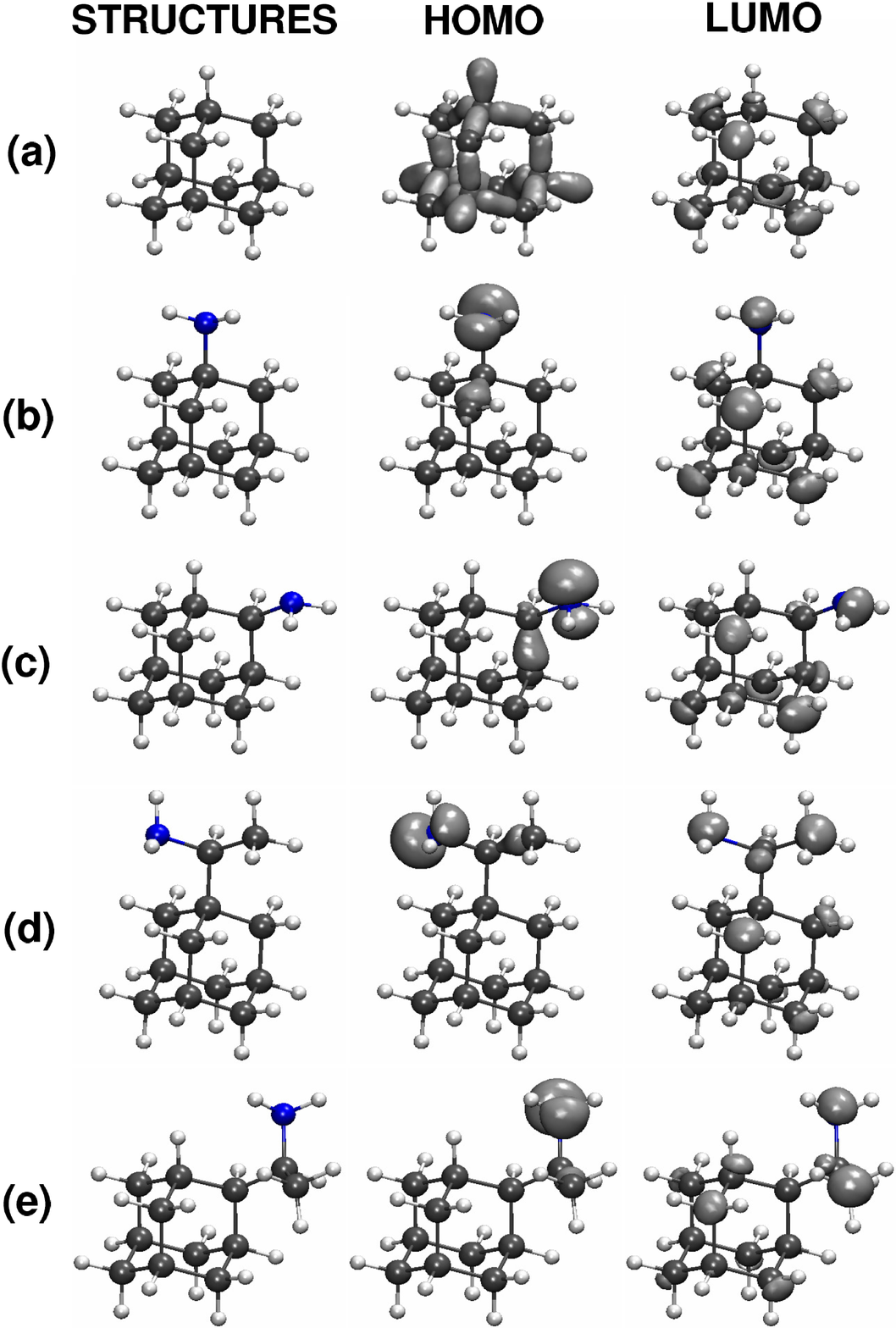}}  
\caption{(color online) Representation (left panel) of the optimized molecular  
structures of (a) adamantane, (b) 1--amantadine, (c) 2--amantadine, 
(d) 1--rimantadine, and (e) 2--rimantadine. Black, dark gray (blue) and  
light gray spheres represent respectively carbon, nitrogen, and hydrogen atoms. 
The figure also presents the charge isosurface, representing the spatial charge 
distribution of the respective HOMO (middle panel) and LUMO (right panel).}  
\label{fig2}  
\end{figure}  
\end{small}  
\pagebreak

\begin{small}  
\begin{figure}  
\centerline{\includegraphics[width=160mm]{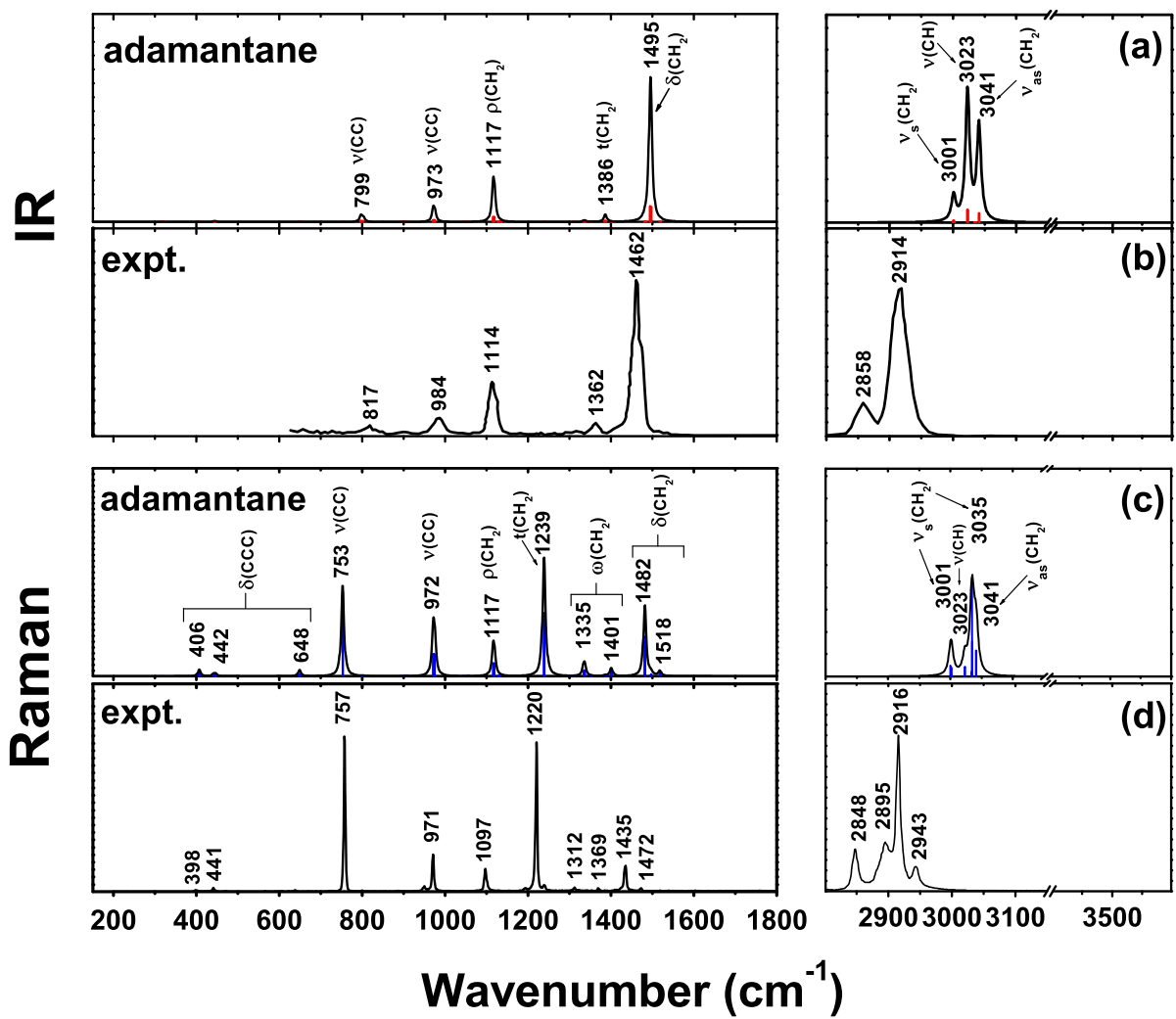}}  
\caption{Vibrational spectra of adamantane: (a) theoretical
and (b) experimental IR spectra \cite{nist}, (c) theoretical 
and (d) experimental  Raman spectra \cite{filik1}.
Activity intensities are given in arbitrary unities (scaled with each other) and 
frequencies in cm$^{-1}$. The modes are labeled by the following symbols:
 $\delta$ (scissor), $\nu$ (stretching), $\rho$ (rocking), $t$ (twisting),
$\omega$ (wagging), and $\tau$ (torsion). Also, $\nu_{\rm s}$ and $\nu_{\rm as}$
represent symmetric and antisymmetric stretching modes.} 
\label{fig3}  
\end{figure}  
\end{small}  
\pagebreak

\begin{small}  
\begin{figure}  
\centerline{\includegraphics[width=160mm]{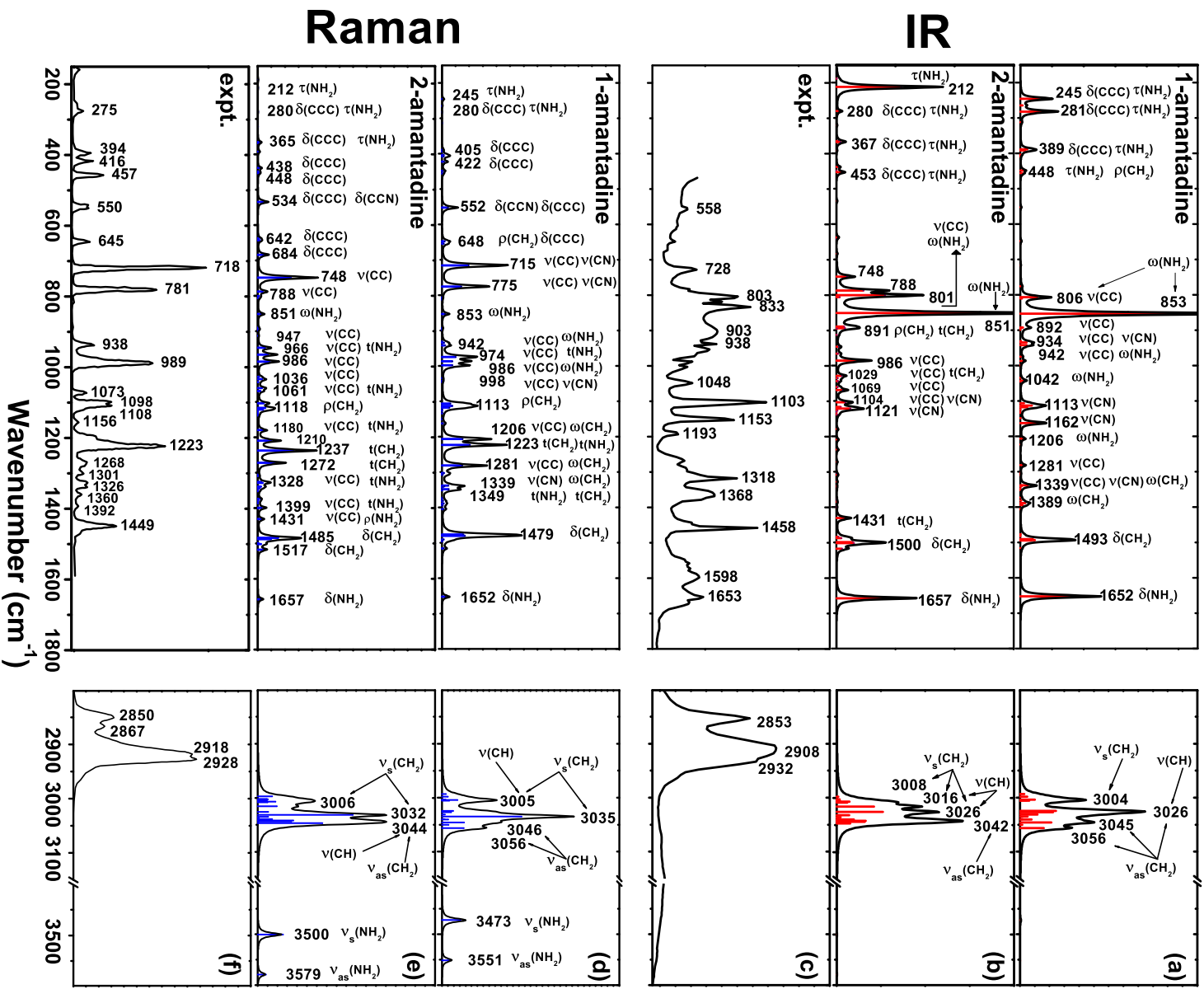}}  
\caption{Vibrational spectra of amantadine isomers: theoretical IR spectra 
of (a) 1--amantadine and (b) 2-amantadine, and (c) experimental IR spectra of 
amantadine samples (in a condensed phase) \cite{nist}. The figure also show the 
theoretical Raman spectra of (d) 1--amantadine and (e) 2-amantadine, and (f) 
experimental Raman spectra of amantadine samples \cite{rivas}.} 
\label{fig4}  
\end{figure}  
\end{small}  
\pagebreak

\begin{small}  
\begin{figure}  
\centerline{\includegraphics[width=160mm]{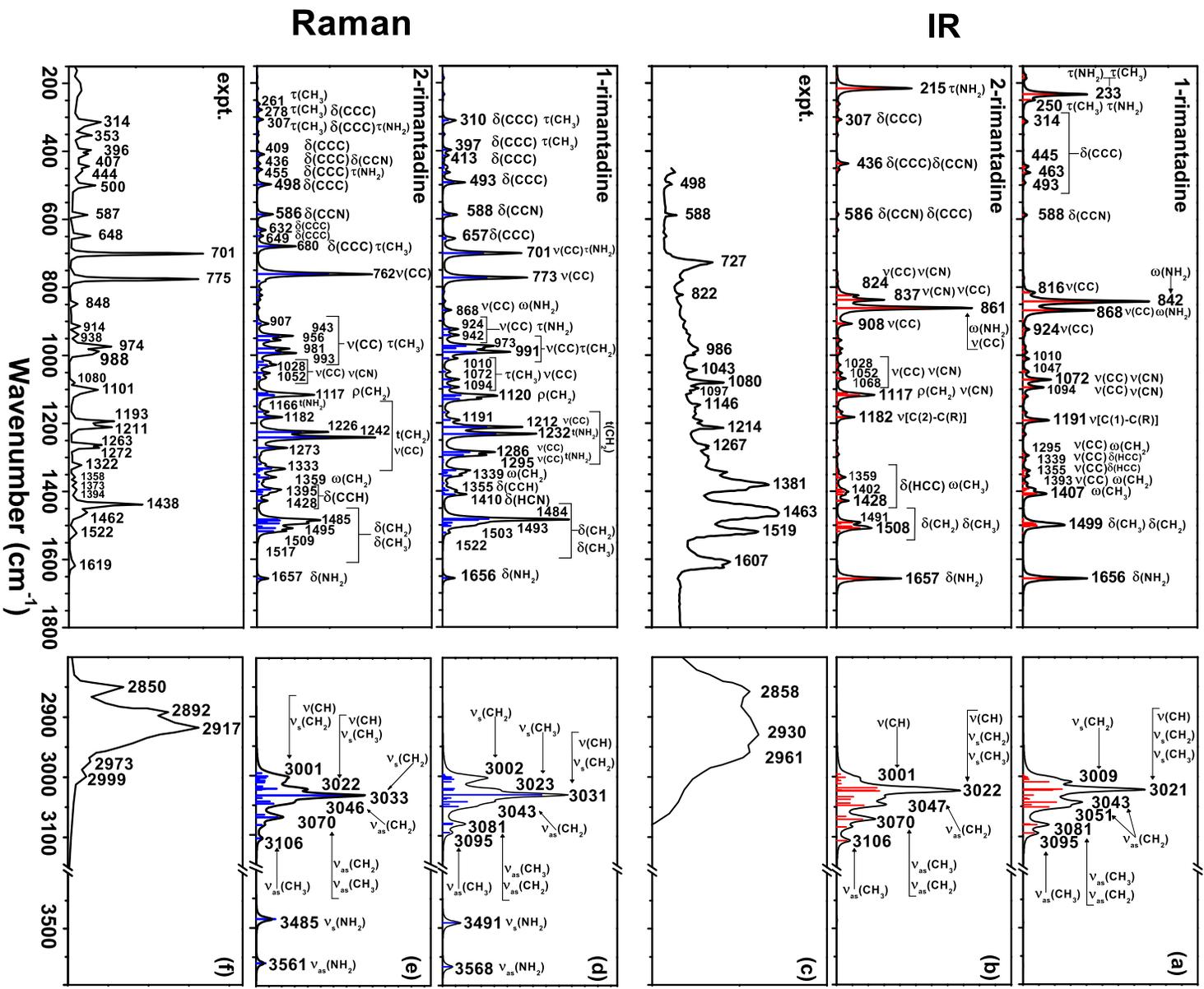}}  
\caption{Vibrational spectra of rimantadine isomers: theoretical IR spectra of (a) 
1--rimantadine and (b) 2-rimantadine, and (c) experimental IR spectra of rimantadine 
samples \cite{sigma}. The figure also show the theoretical Raman spectra of (d) 
1--rimantadine and (e) 2-rimantadine, and (f) experimental Raman spectra of 
rimantadine samples \cite{sigma}.} 
\label{fig5}  
\end{figure}  
\end{small}  
\pagebreak

\end{document}